\documentclass[fleqn,twoside]{article}
\usepackage{amsmath}
\usepackage{espcrc2}
\usepackage{amssymb}
\usepackage{psfrag}
\usepackage{graphicx}
\usepackage{latexsym,amsfonts}
\begin{document}

\title{On the time--modulation of the K--shell electron capture decay
  of H-like ${^{140}}{\rm Pr}^{58+}$ ions produced by neutrino mass
  differences} \author{A. N. Ivanov $^{a,b}$\thanks{\mbox{\it E-mail
  }: ivanov@kph.tuwien.ac.at}, R. Reda$^{b}$,
  P. Kienle$^{b,c}$\thanks{\mbox{\it E-mail }: Paul.Kienle@ph.tum.de},
  \\ \addressmark{$^a$Atominstitut der \"Osterreichischen
  Universit\"aten, Technische Universit\"at Wien, Wiedner Hauptstra\ss
  e 8-10, A-1040 Wien, \"Osterreich, \\ $^b$Stefan Meyer Institut
  f\"ur subatomare Physik, \"Osterreichische Akademie der
  Wissenschaften, Boltzmanngasse 3, A-1090, Wien, \"Osterreich},\\
  $^c$Excellence Cluster Universe Technische Universit\"at M\"unchen,
  D-85748 Garching, Germany}

\date{\today}

\begin{abstract}
  According to recent experimental data at GSI, the rate of the number
  of daughter ions ${^{140}}{\rm Ce}^{58+}$, produced by the nuclear
  K--shell electron capture ($EC$) decay of the H--like ion
  ${^{140}}{\rm Pr}^{58+}$, is modulated in time with a period $T_{EC}
  = 7.06(8)\,{\rm sec}$ and an amplitude $a_{EC} = 0.18(3)$.  We show
  that this phenomenon can be explained by neutrino mass differences
  and derive a value for the difference of squared masses $\Delta
  m^2_{21} = m^2_2 - m^2_1 = 2.22(3)\times 10^{-4}\,{\rm eV}^2$.

\end{abstract}

\maketitle

\subsubsection*{Introduction}
The experimental investigation of the K--shell electron capture ($EC$)
and $\beta^+$ decays of the H--like ${^{140}}{\rm Pr}^{58+}$ and the
He--like ${^{140}}{\rm Pr}^{57+}$ ions, has been recently carried out
in the Experimental Storage Ring (ESR) at GSI in Darmstadt
\cite{GSI1}. The obtained results showed a dependence of the weak
decay rates on the electron structure of the heavy ions. As has been
shown in \cite{Ivanov1}, the experimental data on the ratios of the
weak $EC$ and $\beta^+$--decay rates of the H--like ${^{140}}{\rm
Pr}^{58+}$ and the He--like ${^{140}}{\rm Pr}^{57+}$ ions, obtained in
GSI \cite{GSI1}, can be described within standard theory of weak
interactions of heavy ions and massless Dirac neutrinos \cite{ST2}
with an accuracy better than $3\,\%$.

However, a very recent measurement of the time--dependence of the rate
of the number of daughter ions ${^{140}}{\rm Ce}^{58+}$ from the
$EC$--decay of the H--like ${^{140}}{\rm Pr}^{58+}$ ion, i.e.
${^{140}}{\rm Pr}^{58+} \to {^{140}}{\rm Ce}^{58+} + \nu_e$, showed a
time--modulation of the exponential decay with a period $T_{EC} =
7.06(8)\,{\rm s}$ and an amplitude $a_{EC} = 0.18(3)$
\cite{GSI2}\footnote{For the $EC$--decay of the H--like ${^{142}}{\rm
Pm}^{60+}$ ion the experimental value of the period of the
time--modulation is $T_{EC} = 7.10(22)\,{\rm s}$ with the amplitude of
the time--modulation $a_{EC} = 0.23(4)$ \cite{GSI2}.}. Since the rate
of the number of daughter ions is defined by
\begin{eqnarray}\label{label1}
  \frac{dN^{EC}_d(t)}{dt} = \lambda_{EC}(t)\, N_m(t),
\end{eqnarray}
where $\lambda_{EC}(t)$ is the $EC$--decay rate and $N_m(t)$ is the
number of mother ions ${^{140}}{\rm Pr}^{58+}$, the time--modulation
of $dN^{EC}_d(t)/dt$ implies a periodic time--dependence of the
$EC$--decay rate $\lambda_{EC}(t)$ \cite{GSI2}
\begin{eqnarray}\label{label2}
  \lambda_{EC}(t) = 
\lambda_{EC}\,\Big\{1 + a_{EC}
  \cos\Big(\frac{2\pi t}{T_{EC}} + \phi_{EC}\Big)\Big\},
\end{eqnarray}
where $a_{EC}$, $T_{EC}$ and $\phi_{EC}$ are the amplitude, period and
phase of the time--dependent term \cite{GSI2}.

Nowadays the existence of massive neutrinos, neutrino--flavour mixing
and neutrino oscillations is well established experimentally and
elaborated theoretically \cite{PDG06}. However, these phenomena
concerned mainly the propagation of solar neutrinos and reactor
anti--neutrinos in space and time combined with neutrino oscillations
\cite{PDG06}.  We show that the observed time--modulation of the rate
of the number of daughter ions $dN^{EC}_d(t)/dt$ in the $EC$--decay of
the H--like ${^{140}}{\rm Pr}^{58+}$ ions is not caused by neutrino
oscillations but can be explained by a kind of quantum beats \cite{QB}
due to mass differences of neutrino mass--eigenstates
\cite{PDG06}. This can provide a new method for studying of massive
neutrino mixing, neutrino mass differences and neutrino vacuum
polarisation.
\subsubsection*{Amplitudes of  $EC$--decays of H--like heavy ions}
The Hamilton operator ${\rm H}_W(t)$ of the weak interactions,
responsible for the $EC$ and $\beta^+$ decays of the H--like heavy
ions, can be taken in the standard form \cite{ST2} but accounting for
the neutrino--flavour mixing \cite{PDG06} (see also
\cite{NO1}--\cite{NO3}). This gives $ {\rm H}_W(t ) = \sum_j
U_{ej}{\rm H}^{(j)}_W(t)$, where $U_{ej}$ are matrix elements of the
neutrino mixing matrix $U$ \cite{PDG06}. The Hamilton operator of the
weak interactions ${\rm H}^{(j)}_W(t)$ is defined by
\begin{eqnarray}\label{label3}
\hspace{-0.3in}&&{\rm H}^{(j)}_W(t)
  =\frac{G_F}{\sqrt{2}}V_{ud}\!\!\int\!\!  d^3x
  [\bar{\psi}_n(x)\gamma^{\mu}(1 - g_A\gamma^5) \psi_p(x)]\nonumber\\
\hspace{-0.3in}&& \times [\bar{\psi}_{\nu_j}(x)
  \gamma_{\mu}(1 - \gamma^5)\psi_{e^-}(x)],
\end{eqnarray}
with standard notations \cite{Ivanov1,ST2}.

The massive neutrinos $\nu_j$ in the final state of the $EC$--decay $m
\to d + \nu_e$, where $m$ and $d$ are the mother and daughter ions,
are indistinguishable in principle, since the electron is entangled
with the electron neutrino $\nu_e$ only \cite{NO2}, which is the
superposition of the neutrino mass--eigenstates $|\nu_e\rangle =
\sum_jU^*_{ej}|\nu_j\rangle$ \cite{PDG06}. Such an
indistinguishability of massive neutrinos $\nu_j$ requires to take the
amplitude $A(m \to d + \nu_e)(t)$ of the $EC$--decay as a coherent sum
of the amplitudes $A(m \to d + \nu_j)(t)$ of the $m \to d + \nu_j$
decays, describing three alternative ways of the evolution of the
initial system, i.e. the mother ion $m$ into the daughter ion
$d$. According to Feynman \cite{Feynman}, the {\it probability
amplitude} of the evolution of the quantum system should be taken in
the form of a coherent sum of the {\it probability amplitudes} of
every alternative way of the evolution (see also \cite{Ivanov7}) in
contrast to the recent assertion by Giunti \cite{Giunti2} and Kienert
{\it et al.}  \cite{Lindner}. As a result we get
\begin{eqnarray}\label{label4}
A(m \to d + \nu_e)(t) = \sum_j U_{e j}A(m \to d + \nu_j)(t),
\end{eqnarray}
where the coefficients $U_{ej}$ take into account that the electron
couples to the electron neutrino only. 

In the standard time--dependent perturbation theory \cite{QM1} the
amplitudes $A(m \to d + \nu_j)(t)$, defined in the rest frame of the
mother ion, are given by
\begin{eqnarray}\label{label5}
\hspace{-0.3in} &&A(m \to d + \nu_j)(t) =\nonumber\\
\hspace{-0.3in} &&= - i\int^t_{-\infty}
d\tau\,e^{\textstyle\,\varepsilon\,\tau}\,\langle
d(\vec{q}\,)\nu_j(\vec{k}_j)|H^{(j)}_W(\tau)|m(\vec{0}\,)\rangle,
\end{eqnarray}
where $\vec{q}$ and $\vec{k}_j$ are 3--momenta of the daughter ion and
the neutrino mass--eigenstate $\nu_j$, respectively. For the
regularization of the integral over time we use the $\varepsilon \to
0$ regularization procedure.

For the description of the time modulated interference term in the
$EC$--decay rate of the H--like heavy ion Eq.(\ref{label2}) we assume
a non--conservation of 3--momenta of neutrino mass--eigenstates in
order to deal with differences of 3--momenta of massive neutrinos
playing important role in our analysis of the interference
term. Non--conservation of 3--momenta of massive neutrinos can be
caused by the uncertainties of momenta of the bound electron, proton
and neutron in the elementary $e^- + p \to n + \nu_j$ transition of
the $EC$--decay of the H--like heavy ion (see also \cite{Ivanov4}).

A quantum mechanical description of the $EC$--decays of the H--like
heavy ions with different 3--momenta of neutrino mass--eigenstates can
be carried out using the wave functions of massive neutrinos in the
form of the Gaussian wave packets \cite{WPF}
\begin{eqnarray}\label{label6}
  \hspace{-0.3in}&&\psi_{\nu_j}(\vec{r},t) = (2\pi \delta^2)^{3/2}
\int \frac{d^3k}{(2\pi)^3}\, e^{\,-\,\frac{1}{2}\,\delta^2
(\vec{k} - \vec{k}_j)^2 }\nonumber\\
  \hspace{-0.3in}&&\times\,e^{\textstyle\,i\vec{k}\cdot \vec{r} -
  iE_j(\vec{k}\,)t} u_{\nu_j}(\vec{k},\sigma_{\nu_j}),
\end{eqnarray}
 where $\delta$ is a spatial spread of the massive neutrino $\nu_j$,
$\vec{k}_j$ is the neutrino momentum and $E_j(\vec{k}\,) =
\sqrt{\vec{k}^{\,2} + m^2_j}$ is the energy of a plane wave with the
momentum $\vec{k}$, $u_{\nu_j}(\vec{k},\sigma_{\nu_j})$ is the Dirac
bispinor of the massive neutrino $\nu_j$. In the limit $\delta \to
\infty$ the wave function (\ref{label6}) reduces to a plane wave.

Following \cite{Ivanov1}, we obtain the amplitude of the $EC$--decay
as a function of time $t$
\begin{eqnarray}\label{label7}
  \hspace{-0.3in}&&A(m \to d + \nu_e)(t) = - \sqrt{3} \sqrt{2 M_m 2
  E_d(\vec{q}\,)} \nonumber\\
\hspace{-0.3in}&&\times\,{\cal M}_{\rm GT}\,\langle
  \psi^{(Z)}_{1s}\rangle(2\pi \delta^2)^{3/2}\sum_j U_{ej}
  \sqrt{E_j(\vec{q}\,)}\nonumber\\
   \hspace{-0.3in}&&\times\,e^{\textstyle - \frac{1}{2} \delta^2(\vec{q}
  + \vec{k}_j)^2}\,\frac{\displaystyle e^{\textstyle\,i\,(\Delta
  E_j(\vec{q}\,) - i\,\varepsilon)t}}{ \Delta E_j(\vec{q}\,) -
  i\,\varepsilon}\,\delta_{M_F,-\frac{1}{2}},
\end{eqnarray}
where $\Delta E_j(\vec{q}\,) = E_d(\vec{q}\,) + E_j(\vec{q}\,) - M_m$
is the energy difference of the final and initial state, $~\vec{q}~$
is the 3--momentum of the daughter ion $d$, $E_d(\vec{q}\,) =
\sqrt{\vec{q}^{\,2} + M^2_d}$ and $M_m$ are the energies of the
daughter and mother ions, respectively, and $E_j(\vec{q}\,)$ is the
energy of the neutrino mass--eigenstate $\nu_j$ with momentum
$\vec{q}$, ${\cal M}_{\rm GT}$ is the nuclear matrix element of the
Gamow--Teller transition $m\to d$ and $\langle \psi^{(Z)}_{1s}\rangle$
is the wave function of the bound electron in the H--like heavy ion
$m$, averaged over the nuclear density \cite{Ivanov1}.

The rate of the neutrino spectrum of the $EC$--decay as a function of
time is defined by
\begin{eqnarray}\label{label8}
\hspace{-0.3in}&& \frac{dN_{\nu_e}(t)}{dt} = \frac{1}{2M_m}\int
\frac{d^3q}{(2\pi)^3 2E_d(\vec{q}\,)}\nonumber\\
\hspace{-0.3in}&& \times\frac{1}{2 F + 1}\frac{d}{dt}\sum_{M_F =\pm
\frac{1}{2}}\!\!\!|A(m\to d+\nu_e)(t)|^2.
\end{eqnarray}
The integration over $\vec{q}$ can be carried out if the width
$\delta$ of the wave packets of the wave functions of neutrino
mass--eigenstates is sufficiently large, so that the Gaussian function
$e^{\,- \,\delta^2\,(\vec{q} + \vec{p}\,)^2}$ is localised in the
vicinity of $\vec{q}\simeq -\,\vec{p}$, where $\vec{p} = \vec{k}_j$
for the diagonal term and $\vec{p} = (\vec{k}_i + \vec{k}_j)/2 =
\vec{k}^{(+)}_{ij}$ for the interference term.  The result of the
integration over $\vec{q}$ is given by two terms
\begin{eqnarray}\label{label9}
\frac{dN_{\nu_e}(t)}{dt} = \frac{dN^{(1)}_{\nu_e}(t)}{dt} +
\frac{dN^{(2)}_{\nu_e}(t)}{dt},
\end{eqnarray}
where we have denoted the diagonal term
\begin{eqnarray}\label{label10}
\hspace{-0.3in}&&\frac{dN^{(1)}_{\nu_e}(t)}{dt} = \frac{3}{2F +
  1}\,|{\cal M}_{\rm GT}|^2\, |\langle \psi^{(Z)}_{1s}\rangle|^2\,(\pi
  \delta^2)^{3/2}\nonumber\\
 \hspace{-0.3in} &&\times \sum_jU^*_{ej}U_{ej}\, E_j(\vec{k}_j)\,
  \frac{2\varepsilon }{(\Delta E_j(\vec{k}_j))^2 +
  \varepsilon^2}\,e^{\textstyle \,\varepsilon t}
\end{eqnarray}
and the interference term
\begin{eqnarray}\label{label11}
\hspace{-0.3in}&& \frac{dN^{(2)}_{\nu_e}(t)}{dt} = \frac{3}{2F +
  1}\,|{\cal M}_{\rm GT}|^2\, |\langle \psi^{(Z)}_{1s}\rangle|^2\,(\pi
  \delta^2)^{3/2}\nonumber\\
 \hspace{-0.3in} &&\times\sum_{i > j} U^*_{ei}U_{ej}\,
   e^{\textstyle\,- \delta^2
   (\vec{k}^{(-)}_{ij})^2}\sqrt{E_i(\vec{k}^{(+)}_{ij})
   E_j(\vec{k}^{(+)}_{ij})}\nonumber\\
\hspace{-0.3in} &&\times\,\Big[\frac{2\varepsilon }{(\Delta
E_i(\vec{k}^{(+)}_{ij}))^2 + \varepsilon^2} + \frac{2\varepsilon
}{(\Delta E_j(\vec{k}^{(+)}_{ij} ))^2 +
\varepsilon^2}\Big]\,e^{\textstyle \,\varepsilon t}\nonumber\\
\hspace{-0.3in} &&\times\,\cos\Big[\Big(E_i(\vec{k}^{(+)}_{ij}) -
E_j(\vec{k}^{(+)}_{ij})\Big)t\Big].
\end{eqnarray}
Here $\vec{k}^{(-)}_{ij} = (\vec{k}_i - \vec{k}_j)/2$ and
$\vec{k}^{(+)}_{ij} = (\vec{k}_i + \vec{k}_j)/2$ are the difference
and averaged neutrino momenta and $E_j(\vec{k}^{(+)}_{ij}) =
\sqrt{(\vec{k}^{(+)}_{ij})^2 + m^2_j}$. The former is an approximate
relation between the energy of the massive neutrino $\nu_j$ and a
momentum $\vec{k}^{(+)}_{ij}$, since 3--momenta of massive neutrinos
are not conserved and massive neutrinos are off--shell. As a result
such a relation cannot be used for the analysis of energy differences
$E_i(\vec{k}^{(+)}_{ij}) - E_j(\vec{k}^{(+)}_{ij})$, which are
sensitive to the off--shell and on--shell states of massive
neutrinos. We show this below.
\subsubsection*{The $EC$--decay rate  $\lambda_{EC}$ of
 the H--like ions} The diagonal part of the rate of the neutrino
spectrum $dN^{(1)}_{\nu_e}(t)/dt$ defines the meanvalue of the
$EC$--decay rate $\lambda^{(1)}_{EC} = \lambda_{EC}$. Taking the limit
$\varepsilon \to 0$ in Eq.(\ref{label10}) we get
\begin{eqnarray}\label{label12}
\hspace{-0.3in}&&\frac{dN^{(1)}_{\nu_e}(t)}{dt} = \frac{3}{2F +
  1}\,|{\cal M}_{\rm GT}|^2\, |\langle \psi^{(Z)}_{1s}\rangle|^2\,(\pi
  \delta^2)^{3/2}\nonumber\\
 \hspace{-0.3in} &&\times\sum_j|U_{ej}|^2\,E_j(\vec{k}_j)\,2\pi\,\delta(\Delta
E_j(\vec{k}_j)),
\end{eqnarray}
where the $\delta$--functions $\delta(\Delta E_j(\vec{k}_j))$ with the
arguments $\Delta E_j(\vec{k}_j) = E_d(\vec{k}_j) + E_j(\vec{k}_j) -
M_m$ describe the conservation of energy in the $EC$--decay channels
$m \to d + \nu_j$. The solution of the equations $\Delta
E_j(\vec{k}_j) = 0$ gives the energies of neutrino mass--eigenstates
$\nu_j$ shown in Fig.\,1 with the energy differences $E_i(\vec{k}_i) -
E_j(\vec{k}_j) = \omega_{ij} = \Delta m^2_{ij}/2 M_m$.
\begin{figure}[t]
\centering \includegraphics[width = 0.35\linewidth]{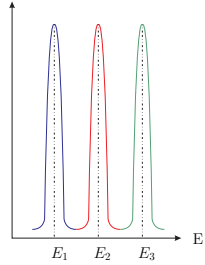}
\caption{Schematic diagram of the neutrino energies $E_j(\vec{k}_j) =
(M^2_m - M^2_d + m^2_j)/2 M_m \simeq Q_{\rm H}$, where $Q_{\rm H} =
M_m - M_d$ is the $Q$--value of the $EC$--decay. The neutrino energy
differences are equal to $E_i(\vec{k}_i) - E_j(\vec{k}_j) =
\omega_{ij} = \Delta m^2_{ij}/2 M_m$.}
\end{figure} 

For the calculation of the $EC$--decay rate $\lambda_{EC}$ we can set
neutrino masses zero and use the unitarity of the $U$--matrix
$\sum_j|U_{ej}|^2 = 1$. This gives
\begin{eqnarray}\label{label13}
  \lambda_{EC} = \int \frac{d^3k}{(2\pi)^3 2E_{\nu}}\, \frac{1}{(\pi
  \delta^2)^{3/2}}\,\frac{dN^{(1)}_{\nu}(t)}{dt},
\end{eqnarray}
where $E_{\nu} = |\vec{k}\,|$ and $(\pi \delta^2)^{3/2}$ is related to
the normalisation of the neutrino wave function\footnote{For
sufficiently large $\delta$ one gets $\int
d^3x\,\psi^{\dagger}_{\vec{q}_j}(\vec{r},t)
\psi_{\vec{k}_j}(\vec{r},t) = (\pi
\delta^2)^{3/2}\,e^{\,-\,\delta^2(\frac{\vec{q}_j -
\vec{k}_j}{2})^2}\,2E_j$. In the limit $\delta \to \infty$ the
Gaussian function behaves as $(\pi
\delta^2)^{3/2}\,e^{\,-\,\delta^2(\frac{\vec{q}_j - \vec{k}_j}{2})^2}
\to (2\pi)^3\, \delta^{(3)}(\vec{q}_j - \vec{k}_j)\,$. This leads to
the standard relativistic covariant normalisation of the neutrino wave
function $\int d^3x\,\psi^{\dagger}_{\vec{q}_j}(\vec{r},t)
\psi_{\vec{k}_j}(\vec{r},t) = (2\pi)^3 2E_j\,\delta^{(3)}(\vec{q}_j -
\vec{k}_j)$. For a finite $\delta$ and $\vec{q}_j = \vec{k}_j$ we get
$\int d^3x\,\psi^{\dagger}_{\vec{k}_j}(\vec{r},t)
\psi_{\vec{k}_j}(\vec{r},t) = (\pi \delta^2)^{3/2}\,2E_j$. }. Having
integrated over the neutrino phase volume we obtain \cite{Ivanov1}
\begin{eqnarray}\label{label14}
\lambda_{EC} = \frac{1}{2 F + 1}\,\frac{3}{2}
 |{\cal M}_{\rm GT}|^2 |\langle \psi^{(Z)}_{1s}\rangle|^2
 \frac{Q^2_{\rm H}}{\pi},
\end{eqnarray}
where $Q_{\rm H} = 3.348(6)\,{\rm MeV}$ for the $EC$--decay
${^{140}}{\rm Pr}^{58+} \to {^{140}}{\rm Ce}^{58+} + \nu_e$
\cite{Ivanov1}. The $EC$--decay rate $\lambda_{EC}$ has been
calculated in \cite{Ivanov1} (see also \cite{Ivanov7}) within standard
theory of weak interactions of heavy ions \cite{ST2}. Since
experimental value of the matrix element $U_{13} =
\sin\theta_{13}\,e^{\,i\,\delta_{CP}}$, where $\delta_{CP}$ is a
CP--violating phase \cite{PDG06}, is very close to zero, we set
$\theta_{13} = 0$ and below deal with two neutrino mass--eigenstates
only. The elements of the mixing matrix are set equal to $U_{e 1} =
\cos\theta_{12}$ and $U_{e 2} = \sin\theta_{12}$ \cite{PDG06}.
\subsubsection*{$EC$--decays of H--like heavy ions as analog of
 quantum beats of atomic transitions} The time--modulation of the
$EC$--decays of the H--like heavy ions bears similarity with quantum
beats of atomic transitions \cite{QB}, since in the $EC$--decays one
deals with the transitions from the initial state $|m\rangle$ to the
final state $|d\,\nu_e\rangle$, where the electron neutrino is the
coherent state of two neutrino mass--eigenstates with the energy
difference equal to $\omega_{21} = \Delta m^2_{21}/2 M_m$.  The time
differential detection of the daughter ions from the $EC$--decays in
the GSI experiments with a time resolution $\tau_d \simeq 1\,{\rm s}$
introduces an energy uncertainty $\delta E_d \sim 2\pi\hbar/\tau_d =
4.14\times 10^{-15}\,{\rm eV}$.  Thus, if $\delta E_d > \omega_{21}$,
following the analogy with quantum beats of atomic transitions
\cite{QB} one should expect a periodic time--dependence of the
$EC$--decay rate with a frequency $\omega_{21}$ and a period $T_{EC}
\sim M_m$.
\subsubsection*{Analysis of the  interference term of  $EC$--decay 
rates of H--like heavy ions} The dependence of the interference term
on the average momentum $\vec{k}^{(+)}_{21} = (\vec{k}_2 +
\vec{k}_1)/2$ can produce an impression that the frequency of the
periodic time--dependence of the interference term is of order of
$\Omega_{21} = \Delta m^2_{21}/2 Q_{\rm H}$ only \cite{Gal}
\begin{eqnarray}\label{label15}
&& \cos((E_2(\vec{k}^{(+)}_{21}) - E_1(\vec{k}^{(+)}_{21}))t) =
\nonumber\\ &&= \cos\Big(\frac{m^2_2 - m^2_1}{E_2(\vec{k}^{(+)}_{21})
+ E_1(\vec{k}^{(+)}_{21})}t\Big)\Longrightarrow\nonumber\\ &&
\Longrightarrow \cos\Big(\frac{m^2_2 - m^2_1}{2 Q_{\rm H}}t\Big) =
\cos(\Omega_{21}t),
\end{eqnarray}
where $E_1(\vec{k}^{(+)}_{21})\simeq E_2(\vec{k}^{(+)}_{21}) \simeq
Q_{\rm H}$ and on--shell relations between the energies and momentum
have been used \cite{Gal}. However, this result, being only partly
correct, leads to the missing of the frequency $\omega_{21}$ due to
the use of on--shell relations between energies and momenta for the
energy difference. The existence of the frequency $\omega_{21}$ can be
shown using non--conservation of neutrino 3--momenta, their
differences and noticing that the behaviour of the interference term
is governed by the Gaussian function
$e^{\,-\,\delta^2(\vec{k}^{(-)}_{21})^2}$. \begin{figure}[t]
\centering \includegraphics[width = 0.35\linewidth]{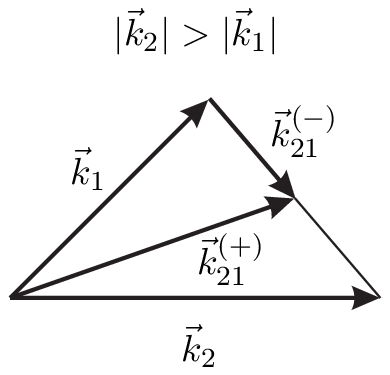}
\includegraphics[width = 0.55\linewidth]{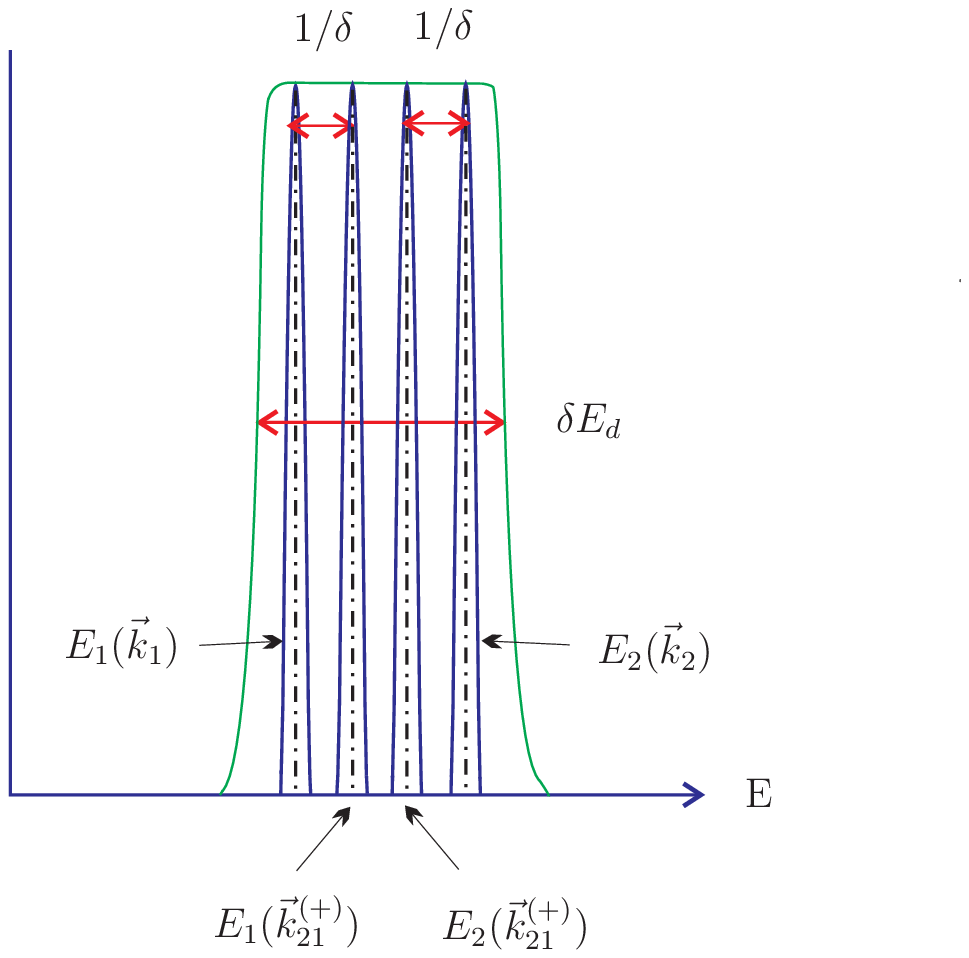}
\caption{For $|\vec{k}_2| > |\vec{k}_1|$ (left) the interference term
  has a periodic time--dependence with a frequency $\omega_{21}$ due
  to the relation $E_1(\vec{k}_1) < E_1(\vec{k}^{(+)}_{21}) <
  E_2(\vec{k}^{(+)}_{21}) < E_2(\vec{k}_2)$ (right) with
  $\vec{k}^{(+)}_{21} = (\vec{k}_2 + \vec{k}_1)/2$.}
\end{figure} 
\begin{figure}[t]
\centering
\includegraphics[width = 0.35\linewidth]{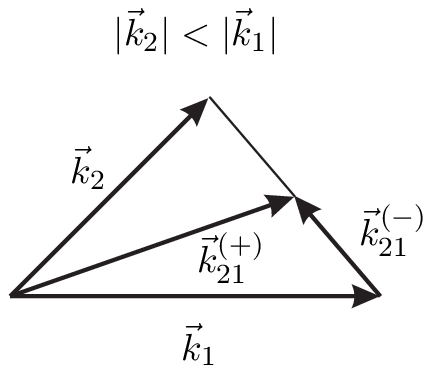}
\includegraphics[width = 0.55\linewidth]{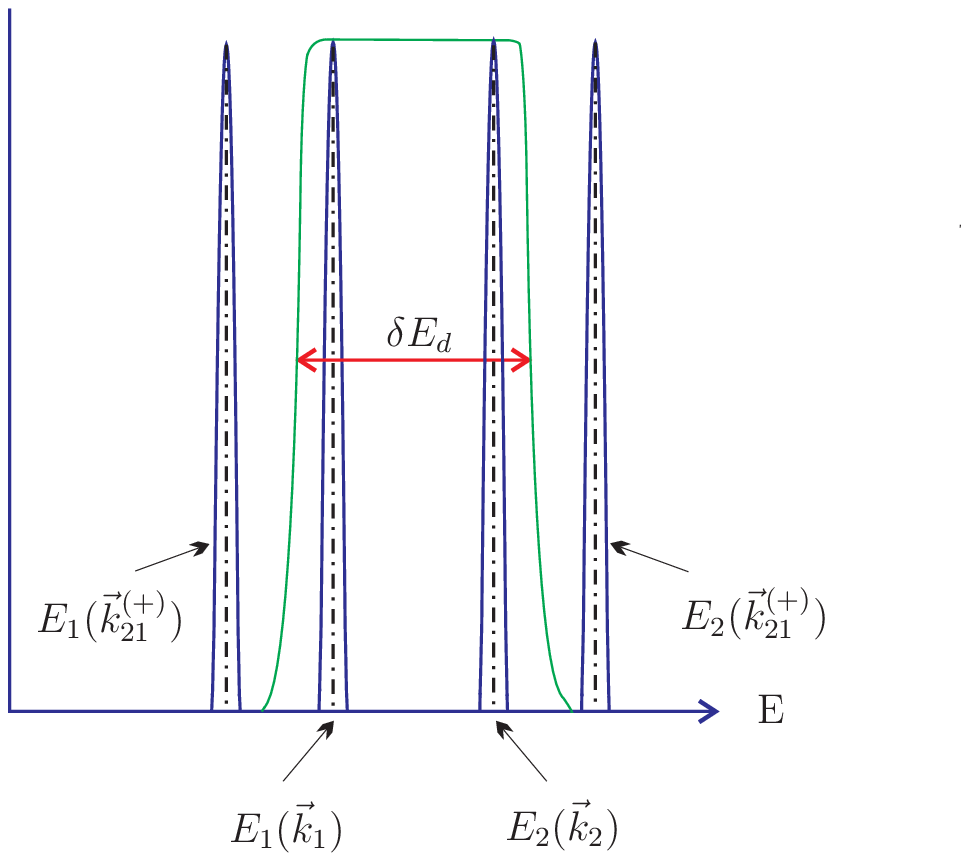}
\caption{For $|\vec{k}_2| < |\vec{k}_1|$ (left) the interference term
  has a periodic time--dependence with a frequency $\Omega_{21} >
  \omega_{21}$ due to the relation $ E_1(\vec{k}^{(+)}_{21}) <
  E_1(\vec{k}_1) < E_2(\vec{k}_2) < E_2(\vec{k}^{(+)}_{21})$ (right)
  with $\vec{k}^{(+)}_{21} = (\vec{k}_2 + \vec{k}_1)/2$.}
\end{figure} 

The region of variation of 3--momenta of massive neutrinos can be
divided into two parts with the absolute values $|\vec{k}_2| >
|\vec{k}_1$ and $|\vec{k}_2| < |\vec{k}_1|$, respectively, and both
cases lead to totally different modulation frequencies of the
interference term.

For $|\vec{k}_2| > |\vec{k}_1$, the energies $E_1(\vec{k}_1)$,
$E_1(\vec{k}_{21})$, $E_2(\vec{k}_{21})$ and $E_2(\vec{k}_2)$ satisfy
the inequality: $E_1(\vec{k}_1) < E_1(\vec{k}_{21}) <
E_2(\vec{k}_{21}) < E_2(\vec{k}_2)$ as it shown in Fig.\,2. The
deviations as indicated in Fig.\,2 are of order of
$O(1/\delta)$. Indeed, for sufficiently large $\delta$, that has been
already assumed above for the calculation the rate of neutrino
spectrum Eq.(\ref{label10}) and Eq.(\ref{label11}), due to the
function $e^{\,-\,\delta^2( \vec{k}^{(-)}_{21})^2}$ the region of the
3--momenta of massive neutrinos is constrained by $\vec{k}_2 \sim
\vec{k}_1$. The allowed region of deviations is of order of
$|\vec{k}^{(-)}_{21}|\sim 1/\delta$. For these momenta the energies of
neutrino mass--eigenstates obey the obvious relation
$E_2(\vec{k}_2)\simeq E_1(\vec{k}_1) \gg |\vec{k}^{(-)}_{21}| \sim
1/\delta$. Making the expansions of the energies of massive neutrinos
in powers of $\vec{k}^{(-)}_{21}$ and keeping only the first order
contributions we get $E_2(\vec{k}^{(+)}_{21}) = E_2(\vec{k}_2) -
\Delta {\cal E}_2$ and $E_1(\vec{k}^{(+)}_{21}) = E_1(\vec{k}_1) +
\Delta {\cal E}_1$, where we have denoted $\Delta {\cal E}_2 =
\vec{k}_2\cdot \vec{k}^{(-)}_{21}/E_2(\vec{k}_2)$ and $\Delta {\cal
E}_1 = \vec{k}_1\cdot \vec{k}^{(-)}_{21}/E_1(\vec{k}_1)$. Without loss
of generality we can set $\Delta {\cal E}_2 \simeq \Delta {\cal E}_1
\simeq \Delta {\cal E}$, where $|\Delta {\cal E}| = O(1/\delta)$ and
$E_2(\vec{k}_2) \simeq E_1(\vec{k}_1) \gg |\Delta {\cal E}| =
O(1/\delta)$. This shows that the deviations of
$E_2(\vec{k}^{(+)}_{21})$ and $ E_1(\vec{k}^{(+)}_{21})$ from
$E_2(\vec{k}_2)$ and $E_1(\vec{k}_1)$, respectively, are of order of
$O(1/\delta)$ and the interference term should have a periodic time
dependence with a period $\omega_{21}$. 

In the region $|\vec{k}_2| >|\vec{k}_1|$ of 3--momenta of massive
neutrinos the calculation of the periodic term runs as follows
\begin{eqnarray}\label{label16}
\hspace{-0.3in}&& \cos((E_2(\vec{k}^{(+)}_{21}) -
E_1(\vec{k}^{(+)}_{21}))t) \stackrel{1/\delta}{\Longrightarrow}
\nonumber\\
\hspace{-0.3in}&& \stackrel{1/\delta}{\Longrightarrow}
\cos((E_2(\vec{k}_2) - E_1(\vec{k}_1))t) = \cos(\omega_{21}t),
\end{eqnarray}
where we have used the on--shell relations between neutrino energies
  and momenta only for the expansions $E_2(\vec{k}^{(+)}_{21}) =
  E_2(\vec{k}_2) - \Delta {\cal E} = E_2(\vec{k}_2) - O(1/\delta)$ and
  $E_1(\vec{k}^{(+)}_{21}) = E_1(\vec{k}_1) + \Delta {\cal E} =
  E_1(\vec{k}_1) + O(1/\delta)$ but not for the energy
  difference. This gives $E_2(\vec{k}^{(+)}_{21}) -
  E_1(\vec{k}^{(+)}_{21}) = E_2(\vec{k}_2) - E_1(\vec{k}_1) - 2 \Delta
  {\cal E} = \omega_{21} - O(2/\delta)$.  As a result the validity of
  the interference term to have a frequency $\omega_{21}$ should be
  confirmed by the inequality $\omega_{21} \gg 2/\delta$, which can be
  always satisfied for the sufficiently large $\delta$.

In turn, for $|\vec{k}_2| < |\vec{k}_1|$ the energies
$E_1(\vec{k}^{(+)}_{21})$, $E_1(\vec{k}_1)$, $E_2(\vec{k}_2)$ and
$E_2(\vec{k}^{(+)}_{21})$ satisfy the inequality:
$E_1(\vec{k}^{(+)}_{21}) < E_1(\vec{k}_1) < E_2(\vec{k}_2) <
E_2(\vec{k}^{(+)}_{21}) $ as it shown in Fig.\,3. In this case a
periodic time--dependence of the interference term can be defined by
the frequency, which is greater than $\omega_{21}$.

In order to take into account both frequencies of the periodic
time-dependence of the interference term we introduce the function
\begin{eqnarray}\label{label17}
\rho(\vec{k}_2,\vec{k}_1) &=& \theta(|\vec{k}_2| -
|\vec{k}_1|)\,\rho'(\vec{k}_2,\vec{k}_1)\nonumber\\ &+&
\theta(|\vec{k}_1| - |\vec{k}_2|)\,\rho''(\vec{k}_2,\vec{k}_1),
\end{eqnarray}
having a meaning of the probability density for neutrino
mass--eigenstates $\nu_2$ and $\nu_1$ to have 3--momenta $\vec{k}_2$
and $\vec{k}_1$, respectively. The Heaviside step functions
$\theta(|\vec{k}_2| - |\vec{k}_1|)$ and $\theta(|\vec{k}_1| -
|\vec{k}_2|)$ take into account two possibilities $|\vec{k}_2| >
|\vec{k}_1|$ and $|\vec{k}_2| < |\vec{k}_1|$ for the 3--momenta of
neutrino mass--eigenstates shown in Fig.\,2 and Fig.\,3.  

The contribution of the interference term $\lambda^{(2)}_{EC}(t)$ to
the $EC$--decay rate we define as follows
\begin{eqnarray}\label{label18}
  &&\lambda^{(2)}_{EC}(t) = \lim_{\varepsilon \to
0}\frac{1}{(\pi \delta^2)^{3/2}}\int
\frac{dN^{(2)}_{\nu_e}(t)}{dt}\nonumber\\
&&\times\,\rho(\vec{k}_2,\vec{k}_1)\,
  \frac{d^3k_1}{(2\pi)^3 2 E_1(\vec{k}_1)}\frac{d^3k_2}{(2\pi)^3 2
  E_2(\vec{k}_2)}.
\end{eqnarray}
Substituting Eq.(\ref{label17}) into
Eq.(\ref{label18}) we get
\begin{eqnarray}\label{label19}
\lambda^{(2)}_{EC}(t) = \lambda^{(2)'}_{EC}(t) +
  \lambda^{(2)''}_{EC}(t),
\end{eqnarray}
where $\lambda^{(2)'}_{EC}(t)$ and $\lambda^{(2)''}_{EC}(t)$ are
determined by the contributions of the regions $|\vec{k}_2| >
|\vec{k}_1|$ and $|\vec{k}_2| < |\vec{k}_1|$, respectively. They are
given by the following momentum integrals
\begin{eqnarray}\label{label20}
\hspace{-0.3in}&&\lambda^{(2)'}_{EC}(t) = \lim_{\varepsilon \to
  0}\frac{3}{2F + 1}\,|{\cal M}_{\rm GT}|^2\, |\langle
  \psi^{(Z)}_{1s}\rangle|^2 \,\frac{1}{2}\,\sin
  2\theta_{12}\nonumber\\
 \hspace{-0.3in} &&\times \int\frac{d^3k_1}{(2\pi)^3 2
  E_1(\vec{k}_1)}\frac{d^3k_2}{(2\pi)^3 2
  E_2(\vec{k}_2)}\,\theta(|\vec{k}_2| - |\vec{k}_1|)\nonumber\\
 \hspace{-0.3in}&&\times\,
  \rho'(\vec{k}_2,\vec{k}_1)\,e^{\textstyle\,- \delta^2
    (\vec{k}^{(-)}_{21})^2}\,\sqrt{E_2(\vec{k}^{(+)}_{21})
  E_1(\vec{k}^{(+)}_{21})}\nonumber\\
\hspace{-0.3in} &&\times\,\Big[\frac{2\varepsilon }{(\Delta
E_2(\vec{k}^{(+)}_{21}))^2 + \varepsilon^2} + \frac{2\varepsilon
}{(\Delta E_1(\vec{k}^{(+)}_{21} ))^2 + \varepsilon^2}\Big]\nonumber\\
\hspace{-0.3in} &&\times\,\cos\Big[\Big(E_2(\vec{k}^{(+)}_{21}) -
E_1(\vec{k}^{(+)}_{21})\Big)t\Big]\,e^{\textstyle \,\varepsilon t}
\end{eqnarray}
and 
\begin{eqnarray}\label{label21}
\hspace{-0.3in}&&\lambda^{(2)''}_{EC}(t) = \lim_{\varepsilon \to
  0}\frac{3}{2F + 1}\,|{\cal M}_{\rm GT}|^2\, |\langle
  \psi^{(Z)}_{1s}\rangle|^2 \,\frac{1}{2}\,\sin
  2\theta_{12}\nonumber\\
 \hspace{-0.3in} &&\times \int\frac{d^3k_1}{(2\pi)^3 2
  E_1(\vec{k}_1)}\frac{d^3k_2}{(2\pi)^3 2
  E_2(\vec{k}_2)}\,\theta(|\vec{k}_1|- |\vec{k}_2|)\nonumber\\
 \hspace{-0.3in}&&\times\,
  \rho''(\vec{k}_2,\vec{k}_1)\,e^{\textstyle\,- \delta^2
  (\vec{k}^{(-)}_{21})^2}\,\sqrt{E_2(\vec{k}^{(+)}_{21})
  E_1(\vec{k}^{(+)}_{21})}\nonumber\\
\hspace{-0.3in} &&\times\,\Big[\frac{2\varepsilon }{(\Delta
E_2(\vec{k}^{(+)}_{21}))^2 + \varepsilon^2} + \frac{2\varepsilon
}{(\Delta E_1(\vec{k}^{(+)}_{21} ))^2 +
\varepsilon^2}\Big]\nonumber\\
\hspace{-0.3in} &&\times\,\cos\Big[\frac{m^2_2 -
m^2_1}{E_2(\vec{k}^{(+)}_{21}) +
E_1(\vec{k}^{(+)}_{21})}t\Big]\,e^{\textstyle \,\varepsilon t}.
\end{eqnarray}
According to Figs.\,2 and 3, the frequencies in two integrals
Eqs.(\ref{label20}) and (\ref{label21}) should be equal to
$\omega_{21}$ and $\Omega_{21}$, respectively. 

Making expansions in powers of $|\vec{k}^{(-)}_{21}| \sim
1/\delta$\,\footnote{Formally this leads to the replacements $\Delta
E_2(\vec{k}^{(+)}_{21}) \to \Delta E_2(\vec{k}_2)$, $\Delta
E_1(\vec{k}^{(+)}_{21}) \to \Delta E_1(\vec{k}_1)$,
$E_2(\vec{k}^{(+)}_{21}) \to E_2(\vec{k}_2)$ and
$E_1(\vec{k}^{(+)}_{21}) \to E_1(\vec{k}_1)$.} and taking the limit
$\varepsilon \to 0$ we reduce the r.h.s. of Eqs.(\ref{label20}) and
(\ref{label21}) to the form
\begin{eqnarray}\label{label22}
  \hspace{-0.3in}&&\lambda^{(2)'}_{EC}(t) = \frac{3}{2F + 1}\,|{\cal
 M}_{\rm GT}|^2\, |\langle \psi^{(Z)}_{1s}\rangle|^2
 \,\frac{1}{2}\,\sin 2\theta_{12}\nonumber\\
 \hspace{-0.3in} &&\times \int\frac{d^3k_1}{(2\pi)^3 2
  E_1(\vec{k}_1)}\frac{d^3k_2}{(2\pi)^3 2
  E_2(\vec{k}_2)}\,\theta(|\vec{k}_2| - |\vec{k}_1|)\nonumber\\
 \hspace{-0.3in}&&\times\,
  \rho'(\vec{k}_2,\vec{k}_1)\,e^{\textstyle\,- \delta^2
  (\vec{k}^{(-)}_{21})^2}\,\sqrt{E_2(\vec{k}_2)
  E_1(\vec{k}_1)}\nonumber\\
\hspace{-0.3in} &&\times\,\Big[2\pi\,\delta(\Delta
E_2(\vec{k}_2)) + 2\pi\,\delta(\Delta
E_1(\vec{k}_1))\Big]\nonumber\\
\hspace{-0.3in} &&\times\cos\Big[\Big(E_2(\vec{k}_2) -
E_1(\vec{k}_1)\Big)t\Big]
\end{eqnarray}
and 
\begin{eqnarray}\label{label23}
\hspace{-0.3in}&&\lambda^{(2)''}_{EC}(t) =  \frac{3}{2F + 1}\,|{\cal
 M}_{\rm GT}|^2\, |\langle \psi^{(Z)}_{1s}\rangle|^2
 \,\frac{1}{2}\,\sin 2\theta_{12}\nonumber\\
 \hspace{-0.3in} &&\times \int\frac{d^3k_1}{(2\pi)^3 2
  E_1(\vec{k}_1)}\frac{d^3k_2}{(2\pi)^3 2
  E_2(\vec{k}_2)}\,\theta(|\vec{k}_1|- |\vec{k}_2|)\nonumber\\
 \hspace{-0.3in}&&\times\,
  \rho''(\vec{k}_2,\vec{k}_1)\,e^{\textstyle\,- \delta^2
  (\vec{k}^{(-)}_{21})^2}\sqrt{E_2(\vec{k}_2)
  E_1(\vec{k}_1)}\nonumber\\
\hspace{-0.3in} &&\times\,\Big[2\pi\,\delta(\Delta E_2(\vec{k}_2)) +
2\pi\,\delta(\Delta E_1(\vec{k}_1))\Big]\nonumber\\
\hspace{-0.3in} &&\times\,\cos\Big[\frac{m^2_2 - m^2_1}{E_2(\vec{k}_2) +
E_1(\vec{k}_1)}\,t\Big].
\end{eqnarray}
By virtue of the $\delta$--functions $\delta(\Delta E_1(\vec{k}_1))$
and $\delta(\Delta E_2(\vec{k}_2))$ with arguments $\Delta
E_1(\vec{k}_1) = E_d(\vec{k}_1) + E_1(\vec{k}_1) - M_m$ and $\Delta
E_2(\vec{k}_2) = E_d(\vec{k}_2) + E_1(\vec{k}_2) - M_m$, respectively,
providing energy conservation in the $EC$--decay channels $m\to d +
\nu_1$ and $m\to d + \nu_2$, and the function $e^{\,- \delta^2
(\vec{k}^{(-)}_{21})^2}$, the frequencies of the periodic terms can be
replaced by $\omega_{21}$ and $\Omega_{21}$, respectively. As a result
the contribution of the interference term reads
\begin{eqnarray}\label{label24}
  \frac{\lambda^{(2)}_{EC}(t)}{\lambda_{EC}} =
a_{EC}\cos(\omega_{21}t) + \tilde{a}_{EC}\cos(\Omega_{21}t),
\end{eqnarray}
where $\lambda_{EC}$ is defined by Eq.(\ref{label14}). The amplitudes
$a_{EC}$ and $\tilde{a}_{EC}$ are given by the integrals
\begin{eqnarray}\label{label25}
  \hspace{-0.3in}&&a_{EC} =\sin 2\theta_{12}\,\frac{2\pi^2}{Q^2_{\rm
  H}}\int\frac{d^3k_1}{(2\pi)^3 2
  E_1(\vec{k}_1)}\frac{d^3k_2}{(2\pi)^3 2
  E_2(\vec{k}_2)}\nonumber\\
 \hspace{-0.3in}&&\times\,
  \rho'(\vec{k}_2,\vec{k}_1)\,e^{\textstyle\,- \delta^2
  (\vec{k}^{(-)}_{21})^2}\sqrt{E_2(\vec{k}_2)
  E_1(\vec{k}_1)}\nonumber\\
\hspace{-0.3in} &&\times\,\Big[\delta(\Delta E_2(\vec{k}_2)) +
\delta(\Delta E_1(\vec{k}_1))\Big]\,\theta(|\vec{k}_2| - |\vec{k}_1|)
\end{eqnarray}
and 
\begin{eqnarray}\label{label26}
\hspace{-0.3in}&&\tilde{a}_{EC} = \sin
  2\theta_{12}\,\frac{2\pi^2}{Q^2_{\rm H}}\int\frac{d^3k_1}{(2\pi)^3 2
  E_1(\vec{k}_1)}\frac{d^3k_2}{(2\pi)^3 2 E_2(\vec{k}_2)}\nonumber\\
 \hspace{-0.3in}&&\times\,
  \rho''(\vec{k}_2,\vec{k}_1)\,e^{\textstyle\,- \delta^2
  (\vec{k}^{(-)}_{21})^2}\sqrt{E_2(\vec{k}_2)
  E_1(\vec{k}_1)}\nonumber\\
\hspace{-0.3in} &&\times\,\Big[\delta(\Delta E_2(\vec{k}_2)) +
\delta(\Delta E_1(\vec{k}_1))\Big]\,\theta(|\vec{k}_1| - |\vec{k}_2|).
\end{eqnarray}
The amplitudes of the periodic functions, depending on the mixing
angle $\theta_{12}$ and the parameter $\delta$, are considered as
empirical parameters determined by the experiment.

We would like to emphasize that due to non--conservation of 3--momenta
the massive neutrinos are off--shell. This means that on--shell
relations $E_j(\vec{p}\,) = (\vec{p}^{\;2} + m^2_j)^{1/2}$ between
energies $E_j(\vec{p}\,)$ and 3-momenta $\vec{p} = \vec{k}_j$ or
$\vec{p} = \vec{k}^{(+)}_{ij}$ are only approximate.  This results in
some constraints on the use of on--shell relations between energies
and momenta. Indeed, for the correct calculation of the periodic
time--dependence of the $EC$--decay rate in the momentum integrals
Eqs.(\ref{label20})--(\ref{label26}) one cannot use on--shell
relations between energies and momenta of massive neutrinos for the
analysis of the differences of neutrino energies and momenta due to
substantial sensitivity of these differences to the off--shell and
on--shell states. As we have shown above (see Eq.(\ref{label15}) and a
discussion below), the use of the on--shell relations between the
energies $E_2(\vec{k}^{(+)}_{21})$ and $E_1(\vec{k}^{(+)}_{21})$ and
the momentum $\vec{k}^{(+)}_{21}$ leads to the missing of the
frequency $\omega_{21}$ (see \cite{Gal}).

\subsubsection*{Time--dependent $EC$--decay rates of H--like heavy ions}
Taking into account the contribution of the interference term, the
total $EC$--decay rate we get in the form
\begin{eqnarray}\label{label27}
 \frac{\lambda_{EC}(t)}{\lambda_{EC}} = 1
  + a_{EC} \cos(\omega_{21}t) + \tilde{a}_{EC}
  \cos(\Omega_{21}t).
\end{eqnarray}
The $EC$--decay rate Eq.(\ref{label27}) contains two periodic
terms. In the laboratory frame the periods are equal to
\begin{eqnarray}\label{label28}
T_{EC} = \frac{2\pi \gamma}{\omega_{21}}\quad,\quad T_g = \frac{2\pi
\gamma}{\Omega_{21}},
\end{eqnarray}
where $\gamma = 1.43$ is the Lorentz factor of the motion of mother
ions in the laboratory frame \cite{GSI2}. We identify $T_{EC} = 2\pi
\gamma/\omega_{21} \sim M_m$ with the experimental period of the
time--modulation $T_{EC} = 7.06(8)\,{\rm s}$, because it is shown
experimentally that the period of the time--modulation does not depend
on the $Q$--value of the weak transition. For the $EC$--decay of the
H--like ${^{142}}{\rm Pm}^{60+}$ ion with $Q_{\rm H} \simeq 4827\,{\rm
keV}$ the period of modulation $T_{EC}= 7.10(22)\,{\rm s}$ is equal to
$T_{EC}= 7.06(8)\,{\rm s}$ of the H--like ${^{140}}{\rm Pr}^{58+}$ ion
within the experimental error bars. Since the mass difference of the
H--like ions ${^{142}}{\rm Pm}^{60+}$ and ${^{140}}{\rm Pr}^{58+}$ is
small we expect only small differences in the periods of the
time--modulation as shown by the experiment.

Thus, for $T_{EC} = 7.06(8)\,{\rm s}$ we get $(\Delta m^2_{21})_{\rm
  GSI} = 2.22(3)\times 10^{-4}\,{\rm eV^2}$ (see also
  \cite{Kleinert}).  The value $(\Delta m^2_{21})_{\rm GSI} =
  2.22(3)\times 10^{-4}\,{\rm eV^2}$ is by a factor of 2.75 larger
  than $(\Delta m^2_{21})_{\rm KamLAND} = 0.80^{+0.06}_{-0.05}\times
  10^{-4}\,{\rm eV^2}$, obtained as a best--fit of the global analysis
  of the solar--neutrino and KamLAND experimental data
  \cite{PDG06,NO3}.

For $(\Delta m^2_{21})_{\rm GSI} = 2.22(3)\times 10^{-4}\,{\rm eV^2}$
we get $T_g \simeq 1.8\times 10^{-4}\,{\rm s}$. These fast
oscillations, averaged over the experimental time resolution $\Delta T
= 0.32\,{\rm s}$ \cite{GSI2}, are not observable. As a result the
time--dependent $EC$--decay rate Eq.(\ref{label27}) is given only by
the low frequency term
\begin{eqnarray}\label{label29}
 \lambda_{EC}(t) = \lambda_{EC}\,(1 + a_{EC}
\cos(\omega_{EC}t)),
\end{eqnarray}
where $\omega_{EC} = \omega_{21} = 2\pi/T_{EC}$ and the amplitude
$a_{EC}$ is defined by Eq.(\ref{label25}).

\subsubsection*{Discussion and summary}
We have shown that the experimental data on the time--modulation
of the rate of the number of daughter ions ${^{140}}{\rm Ce}^{58+}$,
observed in the $EC$--decay of the H--like ion ${^{140}}{\rm
Pr}^{58+}$ \cite{GSI2}, can be explained by the neutrino mass
differences.  However, the difference of squared neutrino masses
$(\Delta m^2_{21})_{\rm GSI} = 2.22(3)\times 10^{-4}\,{\rm eV^2}$,
derived from the period $T_{EC} = 7.06(8)\,{\rm s}$ of the rate of the
number of daughter ions ${^{140}}{\rm Ce}^{58+}$, is 2.75 times larger
than the value measured by KamLAND \cite{PDG06}. A solution of this
problem in terms of neutrino mass corrections, induced by the
interaction of massive neutrinos with a strong Coulomb field of the
daughter ion through virtual $\ell^-W^+$ pair creation, is proposed in
\cite{Ivanov4}.

In summary we argue that the mechanism of the time--dependence of the
$EC$--decay rates of the H--like heavy ions bears similarity with
quantum beats of atomic transitions \cite{QB}, as the $EC$--decays are
the transitions of the initial state $|m\rangle$ into the final state
$|d\nu_e\rangle$, where the electron neutrino is the coherent state of
two massive neutrinos with energy difference $\omega_{21}$. A
sophisticated calculation of the interference term shows also the
existence of a periodic time--dependence with a frequency $\Omega_{21}
\gg \omega_{21}$. In order to take into account the contributions of
two possible frequencies of the periodic time--dependence of the
interference term we have introduced the probability density
$\rho(\vec{k}_2,\vec{k}_1)$ for neutrino mass--eigenstates $\nu_2$ and
$\nu_1$ to get 3--momenta $\vec{k}_2$ and $\vec{k}_1$,
respectively. The calculation of the function
$\rho(\vec{k}_2,\vec{k}_1)$ is rather hard problem. We can only argue
that in the massless limit it should be equal to
\begin{eqnarray*}
\rho(\vec{k}_2,\vec{k}_1) = 2 \sqrt{E_{\nu}(\vec{k}_1)
E_{\nu}(\vec{k}_2)} (2\pi)^3 \delta^{(3)}(\vec{k}_2 -
\vec{k}_1),
\end{eqnarray*}
which is required by the necessity to retain the meanvalue of the
$EC$--decay rate $\lambda_{EC}$.

The necessary condition for the appearance of the interference term in
the $EC$--decay rate is the overlap of the energy levels of neutrino
mass--eigenstates.  Non--conservation of 3--momenta of neutrino
mass--eigenstates provides a possibility to place the interference
term of the $EC$--decay rate in the energy region of the diagonal
term, defining the meanvalue of the $EC$--decay rate $\lambda_{EC}$. A
validity of such a transformation is supported by the constraint $
\omega_{21} = \Delta m^2_{21}/2 M_m \gg 2/\delta$, which makes
impossible the limit $\Delta m^2_{21} \to 0$ in the argument of the
interference term, where the terms of order $O(1/\delta)$ are dropped
with respect to $ \Delta m^2_{21}/4 M_m$. This means that in the
interference term of the $EC$--decay rate Eq.(\ref{label29}) one
cannot set $\Delta m^2_{21}\to 0$ in order to reduce the
time--dependent $EC$--decay rate to the $EC$--decay rate
$\lambda_{EC}$ given by Eq.(\ref{label14}). A correct reduction of
$\lambda_{EC}(t)$ to $\lambda_{EC}$ can be carried out only by means
of the averaging over time $\langle \lambda_{EC}(t)\rangle =
\lambda_{EC}$. In spite of momentum non--conservation, energy of the
$EC$--decay is conserved in our approach.  This is required by {\it
Fermi Golden Rule} \cite{QM1}.

We acknowledge many fruitful discussions with T. Ericson, M. Faber,
 M. Kleber, F. Bosch, Yu. A.  Litvinov, H. Lipkin, E. L. Kryshen,
 M. Pitschmann, N. I. Troitskaya and K. Yazaki in the course of this
 work. We are greatful to M. Gell--Mann and W. Weise for discussions
 and encouraging comments. We thank H. Kleinert for the recommendation
 to use the $\varepsilon \to 0$ regularization for the calculation of
 the integral over time in Eq.(\ref{label5}) and A. Gal for calling
 our attention to the period $T_g$.

This research was partly supported by the DFG cluster of excellence
"Origin and Structure of the Universe".


\begin{thebibliography}{9}
\bibitem{GSI1} 
Yu. A. Litvinov {\it et al.} (the GSI Collaboration),
Phys. Rev. Lett. {\bf 99}, 262501 (2008).
\bibitem{Ivanov1}
A. N. Ivanov {\it et al.},
Phys. Rev. C {\bf 78}, 025503 (2008); arXiv: 0711.3184 [nucl--th].
\bibitem{ST2}
W. Bambynek {\it et al.},
Rev. Mod. Phys. {\bf 49}  (1977) 77.
\bibitem{GSI2} 
Yu. A. Litvinov {\it et al.} (the GSI Collaboration), Phys. Lett. B
{\bf 664}, 162 (2008).
\bibitem{PDG06} W.--M. Yao {\it et al.}, J. Phys. G {\bf 33}, 1
(2006).
\bibitem{QB}
W. W. Chow {\it et al.}
Phys. Rev. A {\bf 11}, 1380 (1975).
\bibitem{NO1}
L. Stodolsky,
Phys. Rev. D {\bf 58}, 036006 (1998);
H. Lipkin,
Phys. Lett. B {\bf 642}, 366 (2006).
\bibitem{NO2}
W. Grimus,\,Lect.\,Notes\,Phys.\,{\bf 629},\,169\,(2004).
\bibitem{NO3}
G. Fogli {\it et al.},
Prog. Part. Nucl. Phys. {\bf 57}, 71, 742 (2006).
\bibitem{Feynman}
R. P. Feynman {\it et al.},
in {\it The Feynman lectures on physics, Quantum mechanics},
Addison--Wesley, Massachusetts, 1966.
\bibitem{Ivanov7} 
A. N. Ivanov {\it et al.},
arXiv: 0807.2750 [nucl--th].
\bibitem{Giunti2}
C. Giunti, 
Phys. Lett. B {\bf 665}, 92 (2005).
\bibitem{Lindner} 
H. Kienert {\it et al.}, arXiv:
0808.2389 [hep--ph].
\bibitem{QM1}
W. Greiner,
in {\it Quantum mechanics, an Introduction},
Springer--Verlag, Berlin, 2001.
\bibitem{Ivanov4} 
A. N. Ivanov {\it et al.}, arXiv: 0804.1311 [nucl--th].
\bibitem{WPF}
C. Giunti, C. W. Kim,
Phys. Rev. C {\bf 58}, 017301 (1998).
\bibitem{Gal}
A. Gal, arXiv: 0809.1213 [nucl--th].
\bibitem{Kleinert}
H. Kleinert, P. Kienle, arXiv: 0803.2938 [nucl--th];
H. Lipkin, arXiv: 0805.0435 [hep-ph].
\end{thebibliography}
\end{document}